\documentclass[12pt]{article}

\usepackage[latin1]{inputenc}
\usepackage{amsmath}
\usepackage{amsfonts}
\usepackage{amssymb,MnSymbol,slashed}
\usepackage{amsxtra}
\usepackage[margin=3cm]{geometry}
\usepackage{color}
\usepackage{hyperref}
\hypersetup{linktocpage=true}
\usepackage{graphicx}
\usepackage{placeins}
\usepackage{upgreek} 
\usepackage{mathrsfs}
\usepackage{accents}
\usepackage{multirow} 
\usepackage{tikz}

\numberwithin{equation}{section}
\newcommand{\beq}{\begin{equation}}
\newcommand{\eeq}{\end{equation}}
\def\be {\begin{equation}}
\def\ee {\end{equation}}
\def\bs#1\es{\begin{split}#1\end{split}}
\def\ba#1\ea{\begin{align}#1\end{align}}
\def\baed#1\eaed{\begin{aligned}#1\end{aligned}}
\def\bged#1\eged{\begin{gathered}#1\end{gathered}}
\def\bea{\begin{eqnarray}}
\def\eea{\end{eqnarray}}

\def\nn{\nonumber}

\def\c{\chi}

\def\h{\eta}

\def\O{\Omega}

\def\r{\rho}
\def\s{\sigma}

\def\t{\tau}
\def\x{\xi}

\newcommand{\cC}{\mathcal{C}}

\newcommand{\cI}{\mathcal{I}}

\def\cO{{{\mathcal O}}}
\def\cM{\mathcal{M}} 
\def\cN{\mathcal{N}}

\def\bP{\mathbb{P}}
\def\bZ{\mathbb{Z}}
\def\bR{\mathbb{R}}
\def\bC{\mathbb{C}}

\def\fr{\frac}

\def\id{\rlap 1\mkern4mu{\rm l}}

\def\ra{\rightarrow}

\newcommand{\til}[1]{ {\tilde{#1}} }

\let\foo\bar 
\renewcommand{\bar}[1]{ {\foo{  #1} }{} }

\newlength{\dhatheight}

\begin{document}

\baselineskip=16pt
\setlength{\parskip}{6pt}

\begin{titlepage}
\begin{flushright}
\parbox[t]{1.4in}{
\flushright MPP-2013-251}
\end{flushright}

\begin{center}

\vspace*{1.5cm}

{\Large \bf  F-Theory on Spin(7) Manifolds:  \\[.3cm] {Weak-Coupling Limit}} 

\vskip 1.5cm

\renewcommand{\thefootnote}{}

\begin{center}
 \normalsize \bf{Federico Bonetti ${}^1$, Thomas W.~Grimm ${}^1$, Eran Palti ${}^2$, Tom G.~Pugh ${}^1$}\footnote{\texttt{bonetti,\ grimm,\ pught @mpp.mpg.de, palti@thphys.uni-heidelberg.de}} 
\end{center}
\vskip 0.5cm

 ${}^1$ \emph{ Max Planck Institute for Physics, \\ 
        F\"ohringer Ring 6, 80805 Munich, Germany} \\[0.25cm]
 ${}^2$ \emph{ Institut f\"ur Theoretische Physik, Universitat Heidelberg, \\
       Philosophenweg 19, D-69120 Heidelberg, Germany}\\ [0.25cm]

\end{center}

\vskip 1.5cm
\renewcommand{\thefootnote}{\arabic{footnote}}

\begin{center} {\bf ABSTRACT } \end{center}

F-theory on appropriately fibered Spin(7) holonomy manifolds is defined to arise as the dual of M-theory on the same space in the limit of a shrinking fiber.
A class of Spin(7) orbifolds can be constructed as quotients of elliptically fibered Calabi-Yau fourfolds 
by an anti-holomorphic involution. The F-theory dual then exhibits one macroscopic dimension that 
has the topology of an interval.
In this work we study the weak-coupling limit of a subclass of such constructions and identify the objects that arise in this limit.
On the Type IIB side we find space-time filling O7-planes as well as O5-planes and orbifold five-planes with a $(-1)^{F_L}$ factor localised on the interval boundaries.
These orbifold planes are referred to as X5-planes and are S-dual to a D5-O5 system.
For other involutions exotic O3-planes and X3-planes
on top of a six-dimensional orbifold singularity can appear.
We show that the objects present preserve a mutual supersymmetry of four supercharges in the bulk of the interval and two
supercharges on the boundary. It follows that in the infinite-interval and weak-coupling limit full four-dimensional $\cN=1$
supersymmetry is restored, which on the Type IIA side corresponds to an enhancement of supersymmetry by
winding modes in the vanishing interval limit.

\end{titlepage}

\newpage
\noindent\rule{\textwidth}{.1pt}		
\tableofcontents
\vspace{20pt}
\noindent\rule{\textwidth}{.1pt}

\setcounter{page}{1}
\setlength{\parskip}{9pt}

\section{Introduction}

F-theory on elliptically fibered Calabi-Yau fourfolds has been studied intensively since it was originally 
proposed as a description of Type IIB string theory with varying string coupling \cite{Vafa:1996xn}.
Compactifications of F-theory on Calabi-Yau fourfolds preserve minimal supersymmetry in 
the non-compact four dimensions as a result of the SU(4) holonomy of the internal geometry. 
However, Berger's classification 
of the special holonomy groups of eight-dimensional manifolds \cite{berger} shows that the 
largest possible special holonomy group is actually Spin(7). Accessing F-theory compactifications on such 
Spin(7) holonomy manifolds has been a long standing problem that was originally raised in \cite{Vafa:1996xn},
but has only been addressed recently in \cite{Bonetti:2013fma}. Indeed a simple generalization of the usual F-theory 
setup to backgrounds with four non-compact Minkowski directions times 
the internal Spin(7) geometry leads to immediate difficulties. 

In order to approach F-theory on Spin(7) manifolds one can, however, 
view F-theory as a particular limit of M-theory. Compactifying 
M-theory on a suitably fibered Spin(7) manifold one obtains an 
F-theory setup in the limit of vanishing fibre volume. Recall that 
this duality requires one T-duality when interpreted within Type II string 
theory. 
This procedure allows the four-dimensional effective 
theory to be determined by an appropriate up-lift of 
the three-dimensional M-theory setup \cite{Denef:2008wq,Grimm:2010ks,Bonetti:2013fma}.
Therefore in order to study F-theory on Spin(7) manifolds we must understand M-theory 
on these spaces and implement the decompactification limit. 
In this paper we investigate these questions using the geometries introduced in \cite{Bonetti:2013fma}. 
Inspired by the work of \cite{Joyce:1999nk}, these are formed by quotienting 
a Calabi-Yau fourfold by an anti-holomorphic and isometric involution $\sigma$. In particular, we 
choose the underlying Calabi-Yau manifold to be elliptically fibered with base $B_3$ and require 
the action of $\sigma$ to be compatible with the fibration. We introduce these geometries in more detail in section \ref{Geoms}. 
It was argued in \cite{Bonetti:2013fma} that the duality of M-theory to F-theory on such Spin(7) manifolds suggests 
that the four macroscopically large directions have boundaries. 
In fact, the additional dimension that grows in the M-theory to F-theory limit may be considered to be an interval. 

An important aspect of these compactifications is the amount of supersymmetry that is preserved. 
Compactifying M-theory on a Spin(7) manifold preserves two real supercharges in three dimensions \cite{Papadopoulos:1995da}, 
which could be heuristically interpreted as $\cN=1/2$ four-dimensional supersymmetry. 
One new advance presented in \cite{Bonetti:2013fma} is an understanding of how to reach the four-dimensional limit 
from such compactifications, which involves an interval on the F-theory side. 
For finite interval length the total space preserves two real supercharges, but 
it is important to answer the more specific question:
How much supersymmetry is preserved in the bulk of the interval, how much on the boundary, and 
what is the interplay between the two? Since the duality between M-theory and F-theory acts fibre-wise 
and preserves supersymmetry, 
understanding these aspects can also shed light on the significantly more 
complicated question of how the amount of supersymmetry preserved 
may be modified on the M-theory side in the vanishing fibre limit. 

We will attack this question by studying the weak-coupling limit of these models. 
This is very interesting in itself. Indeed, one of the beautiful aspects of F-theory and M-theory 
is that they are able to describe complicated string theory constructions from 
a purely geometric perspective. The appearance of orientifold planes and D7-branes 
in the weak-coupling limit of F-theory compactifications on Calabi-Yau manifolds 
is well understood as the Sen limit of the geometry \cite{Sen:1996vd,Sen:1997gv}. We will show that the weak-coupling limits of these Spin(7) constructions include more exotic string theory configurations, for example where O7- and O5-planes are present simultaneously together with certain loci which we term X5-planes. 
Such an X5-plane represents the six-dimensional fixed-point locus of an orbifold action 
dressed with an additional factor of $(-1)^{F_L}$, where $F_L$ is the left-moving 
space-time fermion number, as discussed in \cite{Kutasov:1995te,Sen:1996na,Sen:1998rg,Sen:1998ii,Bergman:1998xv,Hellerman:2005ja}.
These configurations arise for Spin(7) constructions based on  involutions that have three-dimensional fixed 
loci in the base and the elliptic fibers over these has fixed lines. We will also study the case where 
the fixed locus in the base is one-dimensional. In this situation we encounter O3- and O7-planes 
simultaneously and an interesting class of X3-planes and exotic O3-planes confined on 
a six-dimensional orbifold singularity. 
However, yet more exotic 
possibilities exist \cite{Bonetti:2013fma} since the fibers over a fixed point on the base could admit 
a fixed-point free action resulting in a Klein bottle fibre. The analysis of this work 
will not cover these cases. 

Using our results on the weak-coupling limit we are able to sharpen our understanding of the 
supersymmetry properties of these setups. By analysing the weakly coupled planes, their mutually 
preserved supersymmetries, 
as well as aspects such as tadpole cancellation, we will show that in the infinite interval limit 
supersymmetry is enhanced to four supercharges, or $\cN=1$, on the F-theory side. This 
implies that a similar enhancement of supersymmetry must occur on the M-theory side in the 
vanishing fibre limit due to new light winding states. In general this would be a highly non-trivial 
process since it would involve strongly coupled M2-brane winding states becoming light at the singular 
locus of the non-trivial fibration. However, we can avoid these complications by considering Sen's 
weak coupling limit of the underlying Calabi-Yau fourfold geometry. This allows us to approach this 
problem within the framework of perturbative Type II string theory. The relevant winding modes are 
then those of Type IIA string theory on an interval of finite size. The resulting configurations 
can then be more systematically studied by using known approaches to winding 
string states. One can then explicitly check that these states are responsible for 
the enhancement of supersymmetry in the limit of vanishing interval size. 

This work is structured as follows. In section \ref{Geoms} we introduce the relevant Spin(7)
geometries as Calabi-Yau fourfold quotients. We discuss their Sen weak-coupling limit 
and deduce the set of quotients acting on the orientifolded Calabi-Yau threefold that 
emerges. Section \ref{weakcouplingsetups} is devoted to a more detailed analysis of these weak-coupling 
setups. We identify the localized objects and study their supersymmetry properties. 
This allows us to comment on supersymmetry restoration in the large interval limit.

\section{Spin(7) Holonomy Manifolds as Quotients}
\label{Geoms}

In this section we introduce the class of manifolds with special holonomy group Spin(7) (which we will refer to \emph{Spin(7) manifolds} for short) that will be studied in this work.
Recall that a Spin(7) manifold preserves only one covariantly constant nowhere vanishing Majorana-Weyl 
spinor $\eta$. In contrast, a Calabi-Yau fourfold, i.e.~a K\"ahler manifold with SU(4) holonomy, 
has two covariantly constant spinors $\eta_1,\eta_2$. We describe in the following 
how one can construct a Spin(7) manifold starting with a Calabi-Yau fourfold 
and will examine this construction for elliptically fibered Calabi-Yau fourfolds. 
The discussion extends the results already presented in 
\cite{Bonetti:2013fma}, and highlights certain important local properties that 
we will need later. 

\subsection{Generalities on the Quotient Construction} \label{gen_quot}

Let us start with a Calabi-Yau fourfold  $Y_4$ that later on is 
allowed to have certain singularities. We demand that it admits an 
anti-holomorphic and isometric involution $\sigma:Y_4 \rightarrow Y_4$, thus
satisfying
 \ba
    \sigma^2 &= \id \, , &   \sigma^*(g) &= g\, , & \sigma^*(\cI) &= - \cI\, ,
 \ea
 where $g$ and $\cI$ are the metric and complex structure on $Y_4$.
Note that this implies that the K\"ahler form $J$ and (4,0)-form $\Omega$ of $Y_4$ transform 
as
\ba \label{trans-JOmega}
   \sigma^* J &= - J\ , &
   \sigma^* \Omega &= e^{2i\theta} \bar \Omega\ ,
\ea
for some constant $\theta$.
The Spin(7) manifolds under consideration are then constructed 
as quotients 
\beq
    Z_8 = Y_4/\sigma\ . 
\eeq
It is important to stress that in general the manifolds $Z_8$ are singular, with 
a singularity set of even real dimension. Discussing these singularities 
in more detail will be one of the tasks of the remainder of this section. 

Let us next impose that $Y_4$ is an elliptic fibration with 
base $B_3$. This implies that there exists a projection map $\pi : \, Y_4 \rightarrow B_3$
that we demand to be compatible with $\sigma$ and lead to 
a well-defined action $\sigma_B = \sigma|_{B_3}$ on $B_3$.
The elliptic fiber over $B_3$ 
can be described by a Weierstrass equation 
\beq
  y^2 = x^3 + f(u_i)\, x \, z^4 + g(u_i) \, z^6 \ ,
  \label{WeierForm}
\eeq
where $x$, $y$, $z$ are projective coordinates 
in $\mathbb P^2_{2,3,1}$ and $f(u_i)$, $g(u_i)$ are functions of base coordinates $u_i$. 
The base $B_3$ might also be defined by additional polynomial constraints. 
At points of vanishing discriminant 
\beq \label{eq:discriminant}
   \Delta = 4 f^3 + 27 g^2 \ , 
\eeq
the elliptic fiber becomes singular. $\Delta=0$ then defines a complex two-dimensional subspace in $B_3$ 
and determines the location of the space-time filling seven-branes on $B_3$.

Let us denote by $\hat L_{\sigma}$ the fixed-point space of $\sigma$ in $Y_4$.
Its projection to $B_3$ is denoted by $L^B_\sigma = \pi(\hat L_\sigma)$ and 
can equally be obtained as the fixed-point space of $\sigma_B$.
In this work we will consider situations in which the dimension of $L^B_\sigma$ is either one or three. 
The simpler case, which we will call case $(a)$, is when $L^B_\sigma$ is three-dimensional, since in this case the base $B_3$ can be non-singular. 
In a given local patch $U$ on $B_3$ containing a fixed point of $\s_B$ we
can describe the action of $\sigma_B$ in local complex coordinates $(z_1, z_2, z_3)$ 
as
\ba
(a)\qquad (z_1, z_2, z_3) & \ra (\bar z_1, \bar z_2, \bar z_3)\ , & 
&\Rightarrow& 
&\text{$L^B_\sigma(U)$ is three-dimensional.} \label{FixedPointsOnB1}
\ea
A possible alternative that we refer to as case $(b)$ is when $L^B_\sigma$ is one-dimensional.
In this situation $B_3$ cannot be smooth and instead is replaced by an orbifold with 
singularities associated with a discrete group $G$ that contains $\bZ_2$. For simplicity 
we will focus here on the case where $G=\bZ_2$ but the extension to more general 
orbifold singularities may be easily performed. A patch $U$ of $B_3$ near such a 
singularity takes locally the form $\bC^3/\bZ_2$ and may be described locally by 
the complex coordinates $(z_1, z_2, z_3)$ identified by $\rho_{U}: (z_1, z_2, z_3) \rightarrow (-z_1,-z_2,z_3)$. 
The action of $\s_B$ on these coordinates is given by 
\ba
(b)\qquad (z_1, z_2, z_3) & \ra (\bar z_2, - \bar z_1, \bar z_3)\ , & 
& \Rightarrow&
& \text{$L^B_\sigma(U)$ is one-dimensional,}  
\label{FixedPointsOnB2}
\ea
which is an involution on the patch $U$ as $\s_B$ squares to the identification $\r_U$. 

Let us point out two special cases where such a situation occurs. Firstly, one could start 
with a non-singular threefold admitting a global $\bZ_2$ and quotient by this symmetry 
to find the base $B_3$. In fact, this sort of situation naturally arises in toroidal orbifolds. 
Secondly, one may consider the case that $B_3$ is described as a hypersurface or complete intersection 
in a higher-dimensional ambient space exhibiting orbifold singularities as a result 
of scaling identifications. This allows $\sigma_B$ to act as an involution on $B_3$
if it is induced by a symmetry of the ambient space that squares to the identity upon 
using the scalings. Both types of constructions appear in \cite{Joyce:1999nk}\footnote{For a stringy analysis of the Hodge numbers 
of these geometries, see also \cite{Blumenhagen:2001qx}.} 
and alternative Spin(7) constructions also appear in \cite{Cvetic:2001pga,Cvetic:2001zx}.

\subsection{Quotients in the Weak-Coupling Limit} \label{weak_quotients}

The weak-coupling limit of F-theory compactifications was originally discussed in \cite{Sen:1996vd,Sen:1997gv}. In this limit the Weierstrass coefficients $f$ and $g$ appearing in  \eqref{WeierForm} can be expanded as
\ba
f  &= C \h - 3 h^2 \ , & 
g &= h( C \h - 2 h^2 ) + C^2 \c \ . 
\ea
The limit is then given by taking $C\ra 0$ and results in a setup that describes 
O7-planes, which lie at $h = 0$ and D7 branes, which lie at $\h^2 = - 12 h \c$. 

In this weak-coupling limit a quotient associated with the O7-action emerges and this quotient must then be combined with the action of $\s$ in order to determine the full group of symmetries which act on the Calabi-Yau threefold that emerges in the weak-coupling limit. In what follows we will briefly review how this O7-quotient emerges in this limit. 

First let us use the $\bP_{231}$ identification to rescale the torus $z$ coordinate, in \eqref{WeierForm}, to 1. Then 
we note that in the limit as $C\ra 0$ the equation of the torus may then be rewritten in terms 
of the new coordinates $\til x$ and $\til y$, where $x = h \til x$, $y = h^{\fr32} \til y$, as 
\ba
\til y^2 = \til x^3 - 3 \til x - 2 \ ,
\ea 
which is manifestly independent of the base coordinates. The harmonic one form of the torus $\O_1 = \fr{dx}{y}$ 
is given in terms of these rescaled coordinates by $\O_1 = h^{-\fr12} \fr{d\til x}{ \til y}$. The O7-action may then 
be seen by moving once around $h=0$ and noting that $\O_1 \ra - \O_1$. 

The Calabi-Yau threefold which is present in the weak-coupling limit is then the double cover of the base 
such that $\O_1$ becomes single valued. To see this we follow the standard Sen construction 
by adding an additional coordinate $\xi$ along with the polynomial constraint
\beq
  \x^2 = h(u_i)\ ,
\eeq
defining the Calabi-Yau threefold $Y_3$.
The holomorphic orientifold involution is given by
\ba \label{def-sigmah}
   \sigma_{\rm h}:Y_3 \ra Y_3\ , \qquad \xi \ra -\xi\ ,
\ea
 and has O7-planes 
at the fixed points given by $h=0$.
Formally lifting  $\Omega_1$ 
from the base to its double cover $Y_3$
we may then write $\O_1 =\fr{d \til x}{\x \til y}$ 
and see the consistency of the O7-monodromy action
$\O_1 \ra - \O_1$ with the map $\x \ra - \x$. 

Next we can write $\O_1$ as $\O_1 = d Z$ where $Z$ is the complex coordinate of the torus which may 
be expanded in terms of the $A$ and $B$ cycle coordinates $x_A$ and $x_B$ as $Z = x_A + \t x_B$. 
This shows that the action of the holomorphic involution \eqref{def-sigmah} 
induces a reflection $R_{AB}$ of the coordinates of the $A$ and $B$ cycles 
given by $(x_A,x_B) \ra (-x_A,-x_B)$. This formal geometric action on the 
the torus coordinates encodes the intrinsic parities of the Type IIB fields
under the orientifold involution. 

As a further step we study these effects in a setups in which the 
Calabi-Yau fourfold is also quotiented by an anti-holomorphic involution $\s$. 
By considering the action of the different involutions on the ambient space of the 
fiber and demanding the invariance of the polynomial which defines the Calabi-Yau 
fourfold we can deduce the action of $\s$ on the Weierstrass coefficients and the 
functions which appear in the weak-coupling limit. To carry this out explicitly we 
assume that $\s$ acts as 
\ba
\s (f,g,h,\h,\c) = (\bar f, \bar g, \bar h,\bar \h,\bar \c)\ . 
\ea 
We have found this to be the case in all examples we have constructed 
using simple involutions on hyper-surfaces in toric ambient spaces. Then by using that 
\ba
j(\t) = \fr{4 (24 f)^3}{4f^3 + 27 g^2} \ , 
\ea
where $j(\t)$ is the familiar modular invariant $j$-function, we find that 
$\tau(\sigma_B(u_i)) = - \bar \tau(u_i)$ \cite{Bonetti:2013fma}.

We now introduce an anti-holomorphic involution 
\beq \label{def-sigmaah}
  \sigma_{\rm ah} : Y_3 \ra Y_3\ ,
\eeq
induced by $\sigma$.
However, we must note that the action of $\s_B$ on $h$ does not 
uniquely determine the action of $\s_{\rm ah}$ on $\x$ which can either 
be $\x \ra  \bar \x$ or $\x \ra -  \bar \x$. Both choices are related 
by $\sigma_{\rm h}$ given in \eqref{def-sigmah} and 
without loss of generality we can choose $\sigma_{\rm ah}$
to act as $\x \ra  \bar \x$.
As a consequence the action of $\sigma_{\rm ah}$ on the uplift 
of $\Omega_1$ is given by $\O_1 \ra \bar \O_1$.
 Writing $\O_1$ in terms 
of $x_A$ and $x_B$ and combining the action of the two involutions $\sigma_{\rm h}$ and 
$\sigma_{\rm ah}$ on $\O_1$ and $\tau$ we find the corresponding actions $R_{AB},\, R_A$, and $R_{B}$
on the coordinates $(x_A,x_B)$ of the $A$ and $B$ cycles.
The set of combined quotients in the weak limit may then be summarised by
\ba
\sigma_{\rm h} : \  ( u_i , \x ) &\ra ( u_i , - \x )\ ,& R_{AB}:\ (x_A,x_B) &\ra (-x_A,-x_B), \nn \\ 
\s_{\rm ah} :\  ( u_i , \x ) &\ra (\sigma_B( u_i) , \bar \x )\ ,& R_B:\ (x_A,x_B) &\ra (x_A,-x_B) , \nn \\
\s_{\rm h} \sigma_{\rm ah}  :\  ( u_i , \x ) &\ra ( \sigma_B( u_i) , - \bar \x )\ , & R_A:\ (x_A,x_B) &\ra (- x_A,x_B) ,
\label{WeakCouplingQuotients}
\ea
where each line lists the action on $Y_3$ along with the formally induced reflection on an auxiliary $T^2$.
By considering the form of these quotients we see that $\sigma_{\rm h}$ and $\s_{\rm ah}$ always commute on bosons
and that the dimension of the fixed space of  $\s_{\rm ah}$ in $Y_3$ is  
always the same as the dimension of the fixed space of the product $\sigma_{\rm h} \s_{\rm ah}$. 
We note that in the case $(b)$, in which $\sigma_B$ has a one-dimensional fixed space, 
the orbifold singularities of $B_3$ must also be up-lifted to the double cover $Y_3$. 
One can analyze these singularities in local patches analogously to the description
given in section \ref{gen_quot}.

Let us close the section by commenting on the M-theory background that corresponds to the 
weak-coupling limit we have described. Clearly one could compactify M-theory on $Z_8$ 
directly and should recover the above weak-coupling setup as  a specific limit in the geometric 
moduli space. However one may instead follow the prescription above by first going to the 
Sen limit of $Y_4$ and then considering the additional quotient by $\s$. Having done this we 
will then take a further limit in which the M-theory circle becomes small and may then consider 
the set of effective quotients in Type IIA. The local geometry near the fixed points of the 
various involutions can then be analysed separately. 

The holomorphic involution $\s_{\rm h}$ has a four-dimensional fixed space on $Y_3$. Cutting 
out a patch of the two-dimensional space normal to this fixed locus and considering the $T^2$ 
fibers over it we obtain a four-dimensional space that is locally of the form 
\ba
\label{normaltoSigmah}
(S^1_{A} \times S^1_{B} \times \bR^2)/\bZ_2 \ , 
\ea
where $\bR^2$ represents the normal space on $Y_3$ and $S^1_{A}$, $S_{B}^1$ are independent cycles of 
the elliptic fiber such that $S_A^1$ is the M-theory circle and $S_B^1$ is the circle along which 
one applies T-duality to go to F-theory. Let us recall that the geometry of the normal space of 
a lifted O6-plane in M-theory is asymptotically given by $(S^1_A \times \bR^3)/\bZ_2$, 
where $\bZ_2$ inverts all coordinates simultaneously. We may then infer that \eqref{normaltoSigmah} 
signals the presence of an O6-plane localised at a point along the circle $S_B^1$. 
This result is well known and is consistent with the fact that in Type IIB the holomorphic action 
is associated with the presence of O7-planes in the geometry. 

Similarly we can consider the fixed-point sets of the anti-holomorphic involution. In doing this 
we will  focus on case $(a)$ where the fixed space of $\s_B$ is three-dimensional.  It is then 
convenient to combine the actions $\s_{\rm ah}$ and $\s_{\rm h} \s_{\rm ah}$ with the induced 
reflections $R_B$ and $R_A$ to form the products $\s_{\rm ah} R_A $ and $\s_{\rm h} \s_{\rm ah} R_B$. 
The normal space to the fixed-point sets of these total actions is locally given by 
\ba
& (S^1_{B} \times \bR^3)/\bZ_2 \ , &\text{and}&
& (S^1_{A} \times \bR^3)/\bZ_2  \ , 
\ea
respectively. The corresponding Type IIA objects are then given by a six-dimensional orbifold plane 
Orb5 and a O6-plane that wraps the $S_B^1$ cycle. We will comment on this setup in more detail 
in the next section. One can also perform this analysis for the case in which $\s_B$ has a 
one-dimensional fixed space. The objects that arise in this situation will be discussed in 
section \ref{sec:toroidal}.

\section{Weak-Coupling Setups} \label{weakcouplingsetups}

In this section we introduce Type IIB and Type IIA string theory setups 
that can arise in the weak-coupling limit of the geometries introduced in section \ref{Geoms}.
In subsection \ref{weak5planes} we first discuss the case in which the fixed-point locus
of $\sigma_B$ is  three-dimensional, i.e.~the case $(a)$ in \eqref{FixedPointsOnB1}. 
We find that the Type IIB setup contains O5-planes and exotic orbifold five-planes.
The case of a one-dimensional fixed-point set of $\sigma_B$, case $(b)$ in \eqref{FixedPointsOnB2}, is discussed in section \ref{weak3planes}. 
This yields exotic orientifold three-planes and orbifold three-planes that we describe in detail on a torus background. 
In both setups our strategy is to start with a proposed Type IIB setting and then stepwise 
translate the objects which appear into the T-dual Type IIA setting and finally to
the geometry of a Spin(7) manifold. 
That the unusual objects that we have identified preserve mutual supersymmetry in both setups 
can be checked explicitly in torus examples as shown in section \ref{sec:toroidal}.
Collecting these insights we then comment on the supersymmetry 
restoration in the large interval limit in section \ref{susy_restoration}.

\subsection{Weak-Coupling Setup with Five-Planes} \label{weak5planes}

The first setting under consideration is obtained by examining
Type IIB on the background 
\beq \label{IIBbackground}
  \cM^{\rm IIB}_{10}  = (\mathbb{M}^{2,1} \times  S^1 \times Y_3) / G \ , 
\eeq
where  $\mathbb{M}^{2,1} $ is three-dimensional flat space, $Y_3$ is a Calabi-Yau threefold, and
 the symmetry group $G$ is generated by the transformations
 \cite{Bonetti:2013fma} \footnote{We follow the conventions of \cite{Blumenhagen:2013fgp}. }
\ba \label{def-cO12}
   \cO_{1} &= \Omega_p \, \sigma_{\rm h} \, (-1)^{F_L}   \ ,&
   \cO_{2} &= R_3\,  \sigma_{\rm ah} \, (-1)^{F_L} \ .
\ea
The operations $\Omega_p$ and $F_L$ are the world-sheet parity and 
the left-moving space-time fermion number and hence are intrinsically stringy symmetries.
We denote by $R_3$ the reflection of the circle to form an interval $I =  S^1/\mathbb{Z}_2$.
The geometric maps 
$ \sigma_{\rm h}$ and $\sigma_{\rm ah}$ are holomorphic 
and anti-holomorphic involutions of a Calabi-Yau threefold $Y_3$, respectively.
Both are demanded to be isometries and required to commute on bosons, as we discuss in more detail below.
In other words, we 
consider two maps $\sigma_{\rm h / ah}: \ Y_3 \rightarrow Y_3$, $\sigma_{\rm h/ah}^2 = \id$
satisfying
\ba
 \sigma_{\rm h/ah}(\hat g) & = \hat g\ ,&
 \sigma_{\rm h}(\hat I) &= \hat I \ , &
 \sigma_{\rm ah}(\hat I) &=- \hat I\ ,
\ea
where $\hat g$ is the metric on $Y_3$, and $\hat I$ is its complex structure. 
The geometric actions $\sigma_{\rm h}$ and $\sigma_{\rm ah}$
will be identified with the actions introduced in \eqref{def-sigmah} and \eqref{def-sigmaah}.
The complete form of $\cO_1$ and $\cO_2$ was proposed in \cite{Bonetti:2013fma} and 
will be confirmed in the following. 

Since $\sigma_{\rm h}$ is holomorphic its fixed-point set $H_{\sigma_{\rm h}}$ is 
holomorphically embedded in $Y_3$. In order to connect to an F-theory setup we 
will demand in the following that $H_{\sigma_{\rm h}}$ of $\sigma_{\rm h}$ is complex  
two-dimensional. 
This ensures that the fixed points of $\cO_1$ are
O7-planes extending along $\mathbb{M}^{2,1}\times I$ and wrapping $H_{\sigma_{\rm h}}$.
To cancel the tadpoles induced by these negative tension objects the setup should 
also contain D7-branes filling $\mathbb{M}^{2,1}\times I$. The setting obtained by $\cO_1$
is known to arise as the weak-coupling limit of F-theory compactified 
on a Calabi-Yau fourfold \cite{Sen:1996vd,Sen:1997gv}, as we already recalled in section \ref{weak_quotients}.

The action of $\cO_2$ is more unusual as it represents a geometric orbifold action combined with a $(-1)^{F_L}$ action. 
These sorts of exotic orbifolds have been studied in \cite{Kutasov:1995te,Sen:1996na,Sen:1998rg,Sen:1998ii,Bergman:1998xv,Hellerman:2005ja}.
Let us note also that the presence of the reflection $R_3$ is necessary in the $\cO_2$ action, since 
an anti-holomorphic involution $\sigma_{\rm ah}$ alone is a Pin-odd transformation 
and hence would not be a symmetry of the chiral Type IIB string theory. 
In the following we demand that $\sigma_{\rm ah}$ has a real three-dimensional 
fixed-point set $L_{\sigma_{\rm ah}}$. The space $L_{\sigma_{\rm ah}}$
is a special Lagrangian sub-manifold due to the properties of $\sigma_{\rm ah}$.
This implies that the fixed-point set of $\cO_2$ is real six-dimensional 
including the non-compact three-dimensional space-time $\mathbb{M}^{2,1}$. The fixed points of $\cO_2$ 
are located at the ends of the interval $I$. We call the resulting fixed planes
X5-planes and will describe their properties in more detail below.

The geometric actions $\sigma_{\rm h}$ and $\sigma_{\rm ah}$
are required to satisfy the properties
\ba
\sigma_{\rm h}\,   R_3 &= R_3  \, \sigma_{\rm h}  \ , &
\sigma_{\rm ah} \, R_3 &= (-1)^{F_L + F_R} \, R_3 \, \sigma_{\rm ah}  \ , &
\sigma_{\rm h}   \sigma_{\rm ah} &=(-1)^{F_L + F_R} \, \sigma_{\rm ah}   \sigma_{\rm h}  \ ,
\ea
where the factor $(-1)^{F_L + F_R}$ signals commutation on  bosons and
anti-commutation on  ten-dimensional fermions. Under these assumptions one easily
computes the algebra of operators $\cO_1$, $\cO_2$ to be
\ba \label{op_algebra}
\cO_1^2 &= \cO_2^2 = \id  \ , &
\cO_1   \cO_2  &= \cO_2   \cO_1\ .
\ea
Consistently quotienting out by $\cO_1$ and $\cO_2$ implies 
 that  one has to also consider the fixed points of the combined action 
\beq
\cO_3 \equiv   \cO_1   \cO_2   = \Omega_p \, R_3 \, \sigma_{\rm h}  \, \sigma_{\rm ah} \ ,
\eeq
in addition to the O7- and X5-planes introduced above.
The fixed-point loci of this action $\cO_3$ are O5-planes 
that fill $\mathbb{M}^{2,1}$ and wrap the three-dimensional special Lagrangian fixed-point set $L_{\sigma_{\rm h} \sigma_{ \rm ah}}$ of $\sigma_{\rm h} \, \sigma_{\rm ah}$
in $Y_3$. As with the O7-planes, these O5-planes also induce a non-trivial tadpole that has to be cancelled. This requires 
us to include D5-branes into the setup that fill $\mathbb{M}^{2,1}$, localize on $I$ and wrap a three-cycle in $Y_3$ homologous 
to $L_{\sigma_{\rm h} \sigma_{ \rm ah}}$. In the following, we will consider only D5-branes directly wrapping $L_{\sigma_{\rm h} \sigma_{ \rm ah}}$.
A summary of the objects that occur in this setup can be found in table \ref{tab-Oobjects}. 
\begin{table}[h]
\centering
\begin{tabular}{|c|c|c|c|} \hline
symmetry & fixed object & location & tadpoles \\[.1cm] \hline
\rule[-.3cm]{0cm}{.8cm} $\cO_1$ &   O7 & $\mathbb{M}^{2,1} \times I \times H_{\sigma_{\rm h}}$ & add D7 \\
\rule[-.3cm]{0cm}{.8cm} $\cO_2$ & X5 & $\mathbb{M}^{2,1} \times L_{\sigma_{\rm ah}}$ & no tadpole \\
\rule[-.3cm]{0cm}{.8cm} $\cO_3$ & O5 & $\mathbb{M}^{2,1} \times L_{\sigma_{\rm h} \sigma_{ \rm ah}}$ & add D5\\
\hline
\end{tabular}
\begin{minipage}{14cm}
\caption{\small Summary of the symmetry transformations acting on the Type IIB setup \eqref{IIBbackground},
together with the  objects appearing at the associated fixed-point loci, and their location. 
}\label{tab-Oobjects}
\end{minipage}
\end{table}

This implies that the Type IIB weak-coupling limit contains the familiar orientifold planes 
as well as X5-planes. The latter planes have been studied in detail in the 
literature \cite{Kutasov:1995te,Sen:1996na,Sen:1998rg,Sen:1998ii,Bergman:1998xv,Hellerman:2005ja} 
within a different context and given their prominent role it is worthwhile to recall their main features. 
The X5-planes can be seen to be the S-dual of an O5-plane with a single D5-brane on top of it; 
since S-duality maps $(-1)^{F_L} \leftrightarrow \Omega$ in Type IIB we see that the orbifold action 
maps to that of an O5-plane. The presence of the D5-brane on top of it can be inferred from 
tadpole cancellation and the presence of a U(1) symmetry 
supported on the X5-plane which is the S-dual of the gauge symmetry on the D5-brane. 
The U(1) is part of the twisted sector, which is most easily identified in the Type IIA dual that is 
just a simple orbifold as we discuss in more detail below. In fact the local orbifold singularity was studied in a global 
compact setting which is the orbifold limit of a K3 (which is in turn dual to heterotic on $T^4$). 
In this global completion, the U(1) is one of the 16 U(1)s arising from the twisted sector of 
the K3 orbifold limit, or in the geometric regime from dimensionally reducing $C_3$ on one of the 
blow-up cycles and sits in a six-dimensional vector multiplet. 

Having identified the weak-coupling objects in table \ref{tab-Oobjects} we 
now note that they can preserve three-dimensional $\cN=1$ supersymmetry 
along $\mathbb M^{2,1}$. Indeed, compactification on the setup \eqref{IIBbackground}
before performing the quotient with respect to $G$ yields a theory with eight supercharges.
This is reduced to two supercharges by the presence of O7-planes, D7-branes, and X5-planes.
The O7-D7 system does not break supersymmetry completely 
because, in the simple case in which the D7-branes sit on top of the O7-planes, 
all these object wrap the holomorphic cycle $H_{\sigma_{\rm h}}$ in $Y_3$.
In a similar fashion, the X5-plane  and the O5-D5 system do not break supersymmetry completely 
because they wrap special Lagrangian sub-manifolds $L_{\sigma_{\rm ah}}$, $L_{\sigma_{\rm h} \sigma_{\rm ah}}$.
Finally, mutual supersymmetry among these objects can be inferred by noting that
the calibration of the special Lagrangian sub-manifolds is adapted by construction to the complex structure
with respect to which $H_{\sigma_{\rm h}}$ is holomorphic.
We will check mutual supersymmetry explicitly in the case of toroidal models in section \ref{sec:toroidal}.

Let us now   follow the various objects to Type IIA string theory and 
lift them to a geometric Spin(7) setup of F-theory. Firstly, we T-dualize along the $x^3$ 
direction, i.e.~the direction associated to the interval $I=S^1/\mathbb{Z}_2$. 
The resulting Type IIA
background is
\beq \label{IIAbackground}
  \cM_{10}^{\rm IIA} =   (\mathbb{M}^{2,1} \times \tilde S^1 \times Y_3) / \tilde G \ ,
\eeq
where $\tilde S^1$ is the T-dual circle and the symmetry group $\tilde G$ is generated by
the T-duals of $\cO_1$ and $\cO_2$, given  by
 \ba \label{cO-TypeIIA}
\tilde \cO_1   &= \O_p R_3 \, \sigma_{\rm h} (-1)^{F_L}  \ , &
\tilde \cO_2   &= R_3 \, \sigma_{\rm ah} \ ,
\ea
respectively.
We also record the T-dual of the combined action $\cO_3$
\beq
\tilde \cO_3 = \O_p \, \sigma_{\rm h} \, \sigma_{\rm ah} \,  (-1)^{F_L} \ .
\eeq
These expressions for the T-dual actions will be tested in the explicit toroidal model
discussed below.

We  realize that both $\tilde \cO_1$ and $\tilde \cO_3$ are Type IIA orientifold involutions 
that admit O6-planes along their fixed-point loci. 
On the one hand, the O6-planes associated to $\tilde \cO_1$ span $\mathbb{M}^{2,1}$ and wrap the 
four-cycle $H_{\sigma_h}$ in $Y_3$. On the other hand, the O6-planes arising from $\tilde \cO_3$
span $\mathbb{M}^{2,1} \times \tilde I$, where $\tilde I=\tilde S^1/\mathbb{Z}_2$ is the T-dual interval, 
and wrap the three-cycle $L_{\sigma_{\rm ah}}$ and $L_{\sigma_{\rm h} \sigma_{\rm ah}}$. In contrast $\tilde \cO_2$ is simply an 
orbifold action on the compact part of \eqref{IIAbackground}. Its fixed loci are six-dimensional orbifold planes denoted by Orb5. 
The fixed-point objects which appear in  Type IIA are summarised in table \ref{tab-IIAOobjects}.
\FloatBarrier
\begin{table}[h]
\centering
\begin{tabular}{|c|c|c|c|} \hline
symmetry & fixed object & location & tadpoles \\[.1cm] \hline
\rule[-.3cm]{0cm}{.8cm} $\tilde \cO_1$ &   O6 & $\mathbb{M}^{2,1} \ \times H_{\sigma_{\rm h}}$ & add D6 \\
\rule[-.3cm]{0cm}{.8cm} $\tilde \cO_2$ & Orb5 & $\mathbb{M}^{2,1} \times L_{\sigma_{\rm ah}}$ & no tadpole \\
\rule[-.3cm]{0cm}{.8cm} $\tilde \cO_3$ & O6 & $\mathbb{M}^{2,1} \times \tilde I \times L_{\sigma_{\rm h} \sigma_{\rm ah}}$ & add D6 \\
\hline
\end{tabular}
\begin{minipage}{14cm}
\caption{\small Summary of the symmetry transformations acting on the T-dual Type IIA setup \eqref{IIAbackground},
together with the  objects appearing at the associated fixed-point loci, and their location. 
} \label{tab-IIAOobjects}
\end{minipage}
\end{table}
\FloatBarrier   

In order to lift these quotients to M-theory we begin by noting that the parts of the quotients 
which do not act on the IIA geometry arise from the reduction of quotients in M-theory as 
\ba
R_{11} &\ \ra \ \O_p (-1)^{F_L} \ ,& 
\cC &\ \ra\ \O_p \ ,
\label{MtheoryQuotientLifts}
\ea
where $\cC$ maps the M-theory three-form as $C_3 \ra -C_3$. This then implies that 
the quotients  \eqref{cO-TypeIIA} are descended from M-theory quotients which act as 
\ba \label{cOM_1}
\tilde \cO_1^M   &= R_3 R_{11} \, \sigma_{\rm h}  \ , & 
\tilde \cO_2^M   &= R_3 \, \sigma_{\rm ah} \ , &
\tilde \cO_3^M   &= R_{11} \, \sigma_{\rm h} \sigma_{\rm ah} \ .
\ea
Identifying the 11 and 3 directions with the $A$ and $B$ cycles of the elliptic fiber respectively, 
these quotients can then be matched to the quotients appearing in \eqref{WeakCouplingQuotients}. 
 
For many applications, such as checking the 
supersymmetry properties of the setup in section \ref{sec:toroidal}, it turns out to be convenient 
to introduce the configurations on a six-torus $T^6$ instead of $Y_3$.
Real coordinates on the 
ten-dimensional background $\mathbb M^{2,1} \times S^1 \times T^6$ are denoted by $x^m$, $m=0,\ldots ,9$.
In the internal space 
$T^6$ they combine into complex coordinates $z_i$, $i=1,2,3$ as $z_1 = x^4 + i x^5$, $z_2 = x^6 + i x^7$, $z_3 = x^8 + i x^9$. 
We implement the holomorphic involution $\sigma_{\rm h}$
and the anti-holomorphic involution $\sigma_{\rm ah}$  as
\ba
\sigma_{\rm h}: (z_1, z_2, z_3) &\ra (z_1, z_2, -z_3) \ , &
\sigma_{\rm ah}: (z_1, z_2, z_3) &\ra (\bar z_1, \bar z_2, \bar z_3) \ .
\ea
Hence  the actions \eqref{def-cO12} take 
the form
\ba \label{cOinv-R}
\cO_1  &= \O_p \, R_{8 9} \, (-1)^{F_L} \ , &
\cO_2  &= R_{3 5 7 9} \, (-1)^{F_L}\ , &
 \cO_3 &= \Omega_p \, R_{3578}\ ,
\ea
where $R_m$ denotes the reflection of the real coordinate $x^m$, and
$R_{m_1 \dots m_N} = R_{m_1} \dots R_{m_N}$.
This implies that the extended fixed-point objects of $\cO_1$, $\cO_2$, and $\cO_3= \cO_1   \cO_2$ are 
extended along the $x^m$-directions as listed in table \ref{tab-xextOobjects}.
\FloatBarrier   
\begin{table}[h]
\centering
\begin{tabular}{|c|c|ccc|c|cccccc|} \hline
symmetry & fixed object &   $x^0$ & $x^1$ &  $x^2$ &  $x^3$ & $x^4$ & $x^5$ & $x^6$ &$x^7$ &$x^8$ & $x^9$ \\[.1cm] \hline
\rule[-.3cm]{0cm}{.8cm} $\cO_1$ &   O7 & $\times$& $\times$& $\times$& $\times$& $\times$& $\times$& $\times$& $\times$ &&\\
\rule[-.3cm]{0cm}{.8cm} $\cO_2$ & X5 &  $\times$& $\times$& $\times$& & $\times$ &&  $\times$ & & $\times$&\\
\rule[-.3cm]{0cm}{.8cm} $\cO_3 = \cO_1 \, \cO_2$ & O5 &  $\times$ & $\times$& $\times$ &&$\times$ & & $\times$ & && $\times$ \\
\hline
\end{tabular}
\begin{minipage}{14cm}
 \caption{\small The location of the fixed-point sets of the Type IIB involutions \eqref{cOinv-R} are displayed in  coordinates $x^m$
 for the toroidal model on $\mathbb M^{2,1} \times S^1 \times T^6$. The symbol $\times$
             indicates that the object fills this dimension. In all other directions the objects are at fixed points.}\label{tab-xextOobjects}
\end{minipage}
\end{table}
\FloatBarrier   

We can now study the dual Type IIA picture obtained 
by T-duality along $x^3$. 
The background is $\mathbb M^{2,1} \times \tilde S^1 \times T^6$, and the 
actions on this background read
 \ba \label{cOinv-RIIA}
\tilde \cO_1   &= \O_p \,  R_{3 8 9} \,  (-1)^{F_L} \ , &
\tilde \cO_2   &= R_{3 5 7 9}   \ , &
\tilde \cO_3   &= \O_p \,  R_{5 7 8} \,  (-1)^{F_L}  \ .
\ea
In this toroidal model one can evaluate explicitly $\tilde \cO_i = T_3 \cO_i T^{-1}_3$, with $T_3$ being the operator that
implements T-duality along the $x^3$ coordinate,
using the rules collected in appendix \ref{SymAlgAndTDual}.
The fixed-point loci of $\tilde \cO_1$, $\tilde \cO_2$, and $\tilde \cO_3$ extend
along the real coordinates $x^0, x^1, x^2$,$\tilde x^3$,$ x^4, \dots, x^9$
as shown in table  \ref{tab-xextOobjectsIIA}. 

\FloatBarrier
\begin{table}[h]
\centering
\begin{tabular}{|c|c|ccc|c|cccccc|} \hline
symmetry & fixed object &  $x^0$ & $x^1$ &  $x^2$ &  $\tilde x^3$ & $x^4$ & $x^5$ & $x^6$ &$x^7$ &$x^8$ & $x^9$ \\[.1cm] \hline
\rule[-.3cm]{0cm}{.8cm} $\tilde \cO_1$ &   O6  & $\times$& $\times$& $\times$& & $\times$& $\times$& $\times$& $\times$ &&\\
\rule[-.3cm]{0cm}{.8cm} $\tilde \cO_2$ & Orb5   & $\times$& $\times$& $\times$& & $\times$ &&  $\times$ & & $\times$&\\
\rule[-.3cm]{0cm}{.8cm} $\tilde \cO_3 = \tilde \cO_1 \, \tilde \cO_2$ & O6  & $\times$ & $\times$& $\times$ & $\times$ &$\times$ & & $\times$ & && $\times$ \\
\hline
\end{tabular}
\begin{minipage}{14cm}
 \caption{\small The location of the fixed-point sets of the Type IIA involutions \eqref{cOinv-RIIA} are displayed in  coordinates $x^m$
 for the toroidal model on $\mathbb M^{2,1} \times S^1 \times T^6$. The symbol $\times$
             indicates that the object fills this dimension. In all other directions the objects are at fixed points.} 
             \label{tab-xextOobjectsIIA}
\end{minipage}
\end{table}
\FloatBarrier

The M-theory lift of this toroidal Type IIA background is completely analogous 
to the general case discussed in \eqref{cOM_1}. For the convenience of 
the reader we summarize the quotients and objects that lie at the 
fixed spaces in table \ref{tab-summary_actions_(a)}.

\FloatBarrier
\begin{table}[h]
\centering
\begin{tabular}{| r @{ $=$ } l |c| r @{ $=$ } l |c| r @{ $=$ } l |} \hline
\multicolumn{3}{|c|}{Type IIB quotient} & \multicolumn{3}{|c|}{Type IIA quotient} & \multicolumn{2}{c|}{M-theory quotient} \\[.1cm] \hline
\rule[-.3cm]{0cm}{.8cm} 
$\cO_1$ & $\Omega_p R_{89} (-1)^{F_L}$ &   O7 & 
$\til \cO_1$ & $\Omega_p R_{389} (-1)^{F_L}$ &   O6 &
$\s_{\rm h} R_{AB}$ & $R_{38911}$\\
\rule[-.3cm]{0cm}{.8cm} 
$\cO_2$ & $R_{3579}  (-1)^{F_L} $  &   X5 & 
${\til \cO}{}_2$ & $R_{3579} \, $  &   Orb5 &
$\s_{\rm ah} R_B$ & $R_{3579} $ \\
$\cO_1 \, \cO_2$&$ \O_p R_{3578} $ &   O5 & 
$\til \cO_1 \, \til \cO{}_2$&$ \O_p R_{578}  (-1)^{F_L}$ &   O6 & 
$\s_{\rm h} \s_{\rm ah} R_A$ & $R_{57811} $ \\
\hline
\end{tabular}
\begin{minipage}{14cm}
 \caption{\small  Summary of the symmetry transformations modded out in Type IIB, Type IIA and M-theory
 in the case that $\sigma_{B}$ has a three-dimensional fixed space. The individual geometric actions have been introduced 
 in section \ref{weak_quotients}.} 
  \label{tab-summary_actions_(a)}
\end{minipage}
\end{table}
\FloatBarrier

\subsection{Weak-Coupling Setups with Three-Planes} \label{weak3planes}

This section is devoted to the situation in which the fixed-point locus of the anti-holomorphic
involution on the base manifold is one-dimensional. This is described by case $(b)$ as 
shown in  \eqref{FixedPointsOnB2}. In this case the fixed locus of $\s_{\rm ah}$ sits on top of a $\mathbb Z_2$ orbifold singularity of $Y_3$.  
In the following we refrain from a description of such setups for a general
Calabi-Yau threefold, and rather discuss directly the toroidal model.
This allows us to identify the localized objects that appear in the weak-coupling limit
and to study in section \ref{sec:toroidal} their mutual supersymmetry properties in a controlled way.

The Type IIB background we analyse is obtained starting from $\mathbb M^{2,1} \times S^1 \times T^6/\bZ_2 $
and taking the quotient with respect to the symmetry group
generated by the transformation $\cO_1$ defined in \eqref{cOinv-R} and by the new
transformation $\widehat \cO_2$, where 
\ba
\cO_1  &= \O_p \, R_{8 9} \, (-1)^{F_L} \ , &
\widehat \cO_2 &= R_{3579} \, H \, (-1)^{F_L} \ ,
\ea
and where $H$ denotes the holomorphic action
\beq
H: (z_1, z_2, z_3) \ra (z_2, - z_1, z_3) \ .
\eeq
In this toroidal model the patch $U$ described in \eqref{FixedPointsOnB2} is extended to cover the whole of the internal space so that the $(z_1, z_2, z_3)$ coordinates that we describe are identified by $\rho: (z_1, z_2, z_3) \rightarrow (-z_1,-z_2,z_3)$.

The presence of the factor $R_3$ inside $\widehat \cO_2$ gives rise to the interval
$I = S^1/\mathbb Z_2$ exactly as in the previous sections.   
However in this case the action of  
  $\widehat \cO_2 $ is not directly an involution on the $(z_1, z_2, z_3)$ coordinates. Rather the algebra satisfied by $\cO_1$, $\widehat \cO_2$ is given by
\ba
\cO_1^2 &= \id \  , &
\widehat \cO_2^4 &= \id \ , &
\cO_1 \, \widehat \cO_2  &= \widehat \cO_2  \, \cO_1 \ ,
\ea
where the operation $\widehat \cO_2^2$ reproduces the identification $\rho = R_{4567}$.

The full symmetry group acting on the $(z_1, z_2, z_3)$ coordinates of the 
covering $T^6$ then contains the set of transformations given by $\{ \id, \cO_1, \widehat \cO_2 , \widehat  \cO_2 ^2, \widehat  \cO_2 ^3, \cO_1  \, \widehat \cO_2 ,
 \cO_1 \, \widehat \cO_2 ^2, \cO_1 \,\widehat \cO_2 ^3 \}$ with actions summarized, for convenience, in table \ref{tab-summary_actions}.
To each non-trivial element we can associate a localized object, as follows.
\begin{itemize}
\item $\cO_1$: this involution is associated to O7-planes exactly as discussed in the previous section.
\item $\widehat \cO_2$: this transformation contains the factor $(-1)^{F_L}$ and admits a fixed-point locus that is real four-dimensional,
fills $\mathbb M^{2,1}$, and is localized at the endpoints of the interval. We call the associated objects
X3-planes.
\item $\widehat \cO_2^2$: as mentioned above, this is a standard $\mathbb Z_2$ orbifold action. Its fixed-point locus
is six-dimensional, fills $\mathbb M^{2,1}$ and the interval, and will be denoted by Orb5.
\item $\widehat \cO_2^3$: this transformation gives another X3-plane 
that lies on top of the X3-plane associated to $\widehat \cO_2$. These two X3-planes are identified under $\r$.
\item $\cO_1 \, \widehat \cO_2$: this action contains a factor $\Omega_p$
but its geometric part   squares to the identity only up to the $\mathbb Z_2$ orbifold action.
The associated fixed-point locus is four-dimensional, fills $\mathbb M^{2,1}$,
and is localized at the endpoints of the interval. We refer to the associated objects as
XO3-planes.
\item $\cO_1 \, \widehat \cO_2^2$: in this case we have a factor $\Omega_p \, (-1)^{F_L}$
and the geometric action squares to one without invoking  the $\mathbb Z_2$ orbifold. We thus
find standard O3-planes.
\item $\cO_1 \, \widehat \cO_2^3$: this action gives another XO3-plane that is located 
on to of the XO3-plane at the fixed points of $\cO_1 \widehat \cO_2$. These two XO3-planes are identified under $\r$.

\end{itemize}
The fixed spaces of these quotients and the objects that lie at them are summarized in table \ref{tab-X3objects}.
\FloatBarrier
\begin{table}[h]
\centering
\begin{tabular}{|c|c|ccc|c|cccccc|} \hline
symmetry & fixed object &   $x^0$ & $x^1$ &  $x^2$ &  $x^3$ & $x^4$ & $x^5$ & $x^6$ &$x^7$ &$x^8$ & $x^9$ \\[.1cm] \hline
\rule[-.3cm]{0cm}{.8cm} $\cO_1$ &   O7 & $\times$& $\times$& $\times$& $\times$& $\times$& $\times$& $\times$& $\times$ &&\\
\rule[-.3cm]{0cm}{.8cm} $\cO_1 \, \widehat \cO_2^2$ & O3 
&  $\times$& $\times$& $\times$&  $\times$& &&   & &  &\\
\rule[-.3cm]{0cm}{.8cm} $\widehat \cO_2^2$ &   Orb5 & $\times$& $\times$& $\times$& $\times$&  &  &  &   & $\times $& $\times $\\
\rule[-.3cm]{0cm}{.8cm} $\widehat \cO_2 \ \& \ \widehat \cO_2^3$ &   X3 & $\times$& $\times$& $\times$&  &  &  &  &   & $\times$ &\\
\rule[-.3cm]{0cm}{.8cm} $\cO_1 \, \widehat \cO_2 \ \& \ \cO_1 \, \widehat \cO_2^3$ &  
 XO3 & $\times$& $\times$& $\times$&  &  &  &  &  && $\times$ \\
\hline
\end{tabular}
\begin{minipage}{14cm}
 \caption{\small Localized objects in the Type IIB setup with involutions $\cO_1$ and $\widehat \cO_2$
 are displayed in  coordinates $x^m$
 for the toroidal model on $\mathbb M^{2,1} \times S^1 \times T^6$. The symbol $\times$
 indicates that the object fills this dimension. In all other directions the objects are at fixed points.} 
  \label{tab-X3objects}
\end{minipage}
\end{table}
\FloatBarrier

Let us note that the X3-planes encountered here are the analogs 
of the X5-planes of section \ref{weak5planes}, since they arise from an orbifold 
action dressed with an additional $(-1)^{F_L}$-factor. However, the 
X3-planes can only exist if they are confined to lie within 
the Orb5 locus of the $\widehat \cO_2^2$-action. 
A natural conjecture for the S-dual of an X3-plane appears 
to be a system of XO3-planes, as introduced above, with 
suitable localized three-branes to cancel the tadpole.
It would be desirable to study these configurations
in more detail. 

Having described the Type IIB setup we can apply the rules 
of appendix \ref{SymAlgAndTDual} to determine 
the T-duals of all actions listed above. The M-theory 
up-lifts are then inferred by using \eqref{MtheoryQuotientLifts}.
The resulting Type IIA actions
and the objects that lie at their fixed points together with M-theory 
symmetries are summarized 
in table \ref{tab-summary_actions}.
One can then make contact with the discussion 
of section \ref{weak_quotients} by matching the 
$A$ and $B$ cycles with the 11 and 3 directions, respectively. 

\FloatBarrier
\begin{table}[h]
\centering
\begin{tabular}{| r @{ $=$ } l |c| r @{ $=$ } l |c| r @{ $=$ } l |} \hline
\multicolumn{3}{|c|}{Type IIB quotient} & \multicolumn{3}{|c|}{Type IIA quotient} & \multicolumn{2}{c|}{M-theory quotient} \\[.1cm] \hline
\rule[-.3cm]{0cm}{.8cm} 
$\cO_1$ & $\Omega_p R_{89} (-1)^{F_L}$ &   O7 & 
$\til \cO_1$ & $\Omega_p R_{389} (-1)^{F_L}$ &   O6 &
$\s_{\rm h} R_{AB}$ & $R_{38911}$\\
\rule[-.3cm]{0cm}{.8cm}
$\widehat \cO_2^2$ & $ R_{4567} $ &   Orb5 &
$\widehat {\til \cO}{}_2^2$ & $ R_{4567} $ &   Orb5 &
$\r$ & $ R_{4567} $ \\
\rule[-.3cm]{0cm}{.8cm} 
$\cO_1 \, \widehat \cO_2^2$ & $\O_p R_{456789} (-1)^{F_L}$ & O3 &
$\til \cO_1 \, \widehat {\til \cO}{}_2^2$ & $\O_p R_{3456789} (-1)^{F_L}$ & O2 &
$\s_{\rm h} \r R_{AB}$ & $R_{345678911} $\\
\rule[-.3cm]{0cm}{.8cm} 
$\widehat \cO_2$ & $R_{3579} H  (-1)^{F_L}$  &   X3 & 
$\widehat {\til \cO}{}_2$ & $R_{3579} \, H $  &   Orb3 &
$\s_{\rm ah} R_B$ & $R_{3579} H$ \\
\rule[-.3cm]{0cm}{.8cm} 
$\widehat \cO_2^3$ & $R_{3469}  H (-1)^{F_L}$  &   X3 & 
$\widehat {\til \cO}{}_2^3$ & $R_{3469} \, H $  &   Orb3 & 
$\s_{\rm ah} \r R_B$ & $R_{3469} H$ \\
\rule[-.3cm]{0cm}{.8cm}
$\cO_1 \, \widehat \cO_2$&$ \O_p R_{3578} H$ &   XO3 & 
$\til \cO_1 \, \widehat {\til \cO}{}_2$&$ \O_p R_{578}  H (-1)^{F_L}$ &   XO4 & 
$\s_{\rm h} \s_{\rm ah} R_A$ & $R_{57811} H$ \\
\rule[-.3cm]{0cm}{.8cm}
$\cO_1 \, \widehat \cO_2^3$&$ \O_p R_{3468}  H$ &   XO3 & 
$\til \cO_1 \, \widehat {\til \cO}{}_2^3$&$ \O_p R_{468}  H (-1)^{F_L}$ &   XO4 & 
$ \s_{\rm h} \s_{\rm ah} \r R_A$ & $R_{46811} H$ \\
\hline
\end{tabular}
\begin{minipage}{16cm}
 \caption{\small Summary of the symmetry transformations modded out in Type IIB, Type IIA and M-theory
 in the case that $\sigma_{B}$ has a one-dimensional fixed space. The individual geometric actions have been introduced 
 in section \ref{weak_quotients}.} 
  \label{tab-summary_actions}
\end{minipage}
\end{table}
\FloatBarrier

\subsection{Mutual Supersymmetry in Toroidal Setups} \label{sec:toroidal}

This section is devoted to the study of the mutual supersymmetry properties of the localized objects introduced 
in the above sections \ref{weak5planes} and \ref{weak3planes}. 
Our analysis will be simplified by considering the 
torus setups of table \ref{tab-xextOobjects} and table \ref{tab-X3objects}.
As a result, we do not perform any additional
orbifold quotient and we rather let $Y_3$ be a simple six-torus, even though this
implies a bulk sector with 32 real supercharges. These arguments therefore do not prove the supersymmetry 
of the setups with more complicated geometries. However, they do demonstrate that the unusual objects that 
we describe do not automatically break supersymmetry completely either on their own or when combined with the other sorts of fixed objects we consider. 

Let us first study the setup of section \ref{weak5planes}
with weak-coupling objects listed in table \ref{tab-xextOobjects}.
We also expect that these localized objects
do not break supersymmetry completely, since the for any pair of them
the number of different Dirichlet/Neumann directions is a multiple of four.
As a warm-up for the more involved case of section \ref{weak3planes}, 
we discuss a more explicit way to infer that
this setup preserves a finite amount of supersymmetry.
To this end, it is useful to combine 
 the two ten-dimensional supersymmetry parameters 
 into an R-symmetry doublet $\epsilon = (\epsilon_L, \epsilon_R)^{\sf T}$,
where the subscripts $L$, $R$ refer to their world-sheet origin.
Operators $\cO_i$ are represented as elements of the tensor product of
the R-symmetry group with Spin(1,9). One has
\ba \label{spin_reps}
\cO_1 &= i \sigma^2 \otimes \Lambda(R_{89}) \ , &
\cO_2 &= -\sigma^3 \otimes \Lambda(R_{3579}) \ , &
\cO_3 &= i\sigma^2 \otimes \Lambda(R_{3578}) \ ,
\ea
where the $\sigma$'s are Pauli matrices, and $\Lambda(M)$ denotes 
the Spin(1,9) element associated to $M \in SO(1,9)$. Note that 
$\Omega_p$ is realized as $\sigma^1$, while $(-1)^{F_L}$
corresponds to $-\sigma^3$.
Supersymmetry is preserved if a non-vanishing solution $\epsilon$
is found to the equations
\ba 
\cO_1 \, \epsilon &= \epsilon \ , &
\cO_2 \, \epsilon &= \epsilon \ .
\ea
The analogous condition with $\cO_3$ is not independent. These equations
can be studied explicitly recalling that $\Lambda(R_m) = i \Gamma \Gamma_m$
in the light-cone formalism. One indeed finds that the operator
\beq
\lambda_1 (\cO_1 - \id) + \lambda_2(\cO_2 - \id)
\eeq
has a non-trivial kernel  of relative dimension 1/4
for $\lambda_1, \lambda_2 \in \mathbb C$.
Taking into account that $\epsilon_L$, $\epsilon_R$ are Majorana spinors,
we have proved that the toroidal setup under examination preserves 8 real supercharges.
This may then be further broken if the torus is replaced by a Calabi-Yau threefold.
We will see another application of the toroidal formalism next
where the familiar rule about Dirichlet/Neumann directions fails.
Note also that the representation \eqref{spin_reps} 
can be used to check explicitly the algebra \eqref{op_algebra}
on fermionic fields.

With this preparation we can now also analyse the setup introduced in section \ref{weak3planes}.
The mutual supersymmetry properties of the localized objects listed in table \ref{tab-X3objects} can be studied
explicitly by representing the actions of $\cO_1$ and $\widehat \cO_2$ on the ten-dimensional
supersymmetry parameters. 
We do not need to consider all other symmetries since they are generated 
by $\cO_1$ and $\widehat \cO_2$.
The action of $\cO_1$ was given in \eqref{spin_reps}.
The action of $\widehat \cO_2$ reads
\beq
\widehat \cO_2 = - \sigma^3 \otimes \Lambda(R_{3579})\, \Lambda(H) \ ,
\eeq
 where  
\ba
\Lambda(R_{3579}) &= \Gamma_{3579} \ , &
\Lambda(H) &= \frac 12 (\id - \Gamma_{46})(\id - \Gamma_{57}) \ .
\ea 
We can thus study the operator
\beq
\lambda_1 (\cO_1 - \id) + \lambda_2(\widehat \cO_2 - \id)
\eeq
and show straightforwardly that, for $\lambda_1, \lambda_2 \in \mathbb C$,
it has non-trivial kernel of relative dimension 1/8, thus proving that our toroidal setup preserves four
real supercharges. Note that in this setup the
Dirichlet/Neumann direction rule is not applicable, since we have an orbifold action and
the geometric transformations under examination do not just consist of reflections.
Let us stress again that the amount of preserved supersymmetry will decrease 
further when replacing the torus by a Calabi-Yau manifold. It would be interesting 
to investigate the rules for this breaking in this more general situation.

\subsection{Large-Interval Limit and Supersymmetry Restoration} \label{susy_restoration}

In this section we discuss some properties of the Type IIB
setup described above in the limit in which the size of the interval $I$
is sent to infinity. More precisely, 
we focus on the resulting four-dimensional low-energy effective action 
and we argue that, for any observer in the bulk of $I$,
such a theory is indistinguishable from the four-dimensional $\cN = 1$ effective theory
obtained by quotienting Type IIB with respect to $\cO_1$ only.

In order to simplify the discussion we suppose that the quotient
under the action of $G$ generated by $\cO_1$ and $\cO_2$ is performed in two steps.
In particular, we consider first the quotient under $\cO_2$ and later implement $\cO_1$, 
since the later does not affect the following arguments.
We are interested in the dynamics of 
excitations with wavelength much larger than  the typical size of the internal space
parametrized by coordinates $x^4$, \dots, $x^9$.
This size, in turn, is supposed to be large compared to the string scale.
As a result, the only states that become light as the interval $I$ decompactifies
are states with no winding and with non-vanishing Kaluza-Klein mode
along $x^3$ only. 

Such states are conveniently packaged into four-dimensional fields depending on
$x^0$, \dots, $x^3$ and
satisfying
Dirichlet or Neumann boundary conditions at the endpoints of the interval.
More precisely, invariance under $\cO_2$ implies that expansion of the massless fields of Type IIB supergravity onto 
positive and negative cohomologies of $Y_3$ under $\sigma_{\rm ah}$
yields  four-dimensional fields with definite parity under reflection  of $x^3$.
Fields with negative parity satisfy Dirichlet boundary conditions
at the endpoints of the interval and for finite interval size
cannot be accessed in the low-energy theory, because they always carry
at least one unit of Kaluza-Klein momentum along $x^3$.

When the size of the interval becomes much larger than the typical
wavelength of the excitations we want to study, however,
the states associated to four-dimensional fields with Dirichlet boundary conditions
become accessible again to the low-energy dynamics.
This implies that we can excite fluctuations of all
four-dimensional fields, irrespectively of their parity under reflection of $x^3$.\footnote{
Only Neumann fields can have a constant V.E.V., strictly speaking.
For a Dirichlet field the allowed profile with the minimum energy
is of the form $\sin (x^3/r)$, where $\pi r$ is the length of the interval,
and can be considered approximately as a constant V.E.V.  
in a sufficiently small region in the bulk of the interval.
}
We are thus led to argue that in the limit of infinite interval $I$
the low-energy four-dimensional effective action is the same as the one 
that would be obtained without performing the quotient with respect to $\cO_2$. Thus,
in this limit the group $G$ effectively reduces to $\cO_1$ only,
and we have a Calabi-Yau orientifold that yields a four-dimensional $\cN=1$ effective action.

We conclude this section with a short remark about the Type IIA
interpretation. The Kaluza-Klein states that become light in the limit on the Type IIB side
correspond to winding states on the the Type IIA side. Kaluza-Klein states 
of a four-dimensional field with Neumann or Dirichlet boundary conditions
at the endpoint of the interval
have the schematic form
\beq
|\psi, n_3 = N, w_3 = 0 \rangle  \pm | \psi,  n_3 = -N, w_3 = 0 \rangle \  , 
\eeq
respectively. In this expression $n_3$, $w_3$ are the Kaluza-Klein
level and winding in the $x^3$ direction,  $N \in \mathbb Z$, and
  $\psi$ is a shorthand notation for the oscillator
structure of the state. T-duality along $x^3$ maps such a state to 
\beq
|\psi, \tilde n_3 = 0, \tilde w_3 = N \rangle  \pm | \psi, \tilde n_3 = 0, \tilde w_3 = -N \rangle \  , 
\eeq
where $\tilde n_3$, $\tilde w_3$ denote Kaluza-Klein level and winding along the 
T-dual coordinate $\tilde x^3$. 

In the uplift to M-theory it is natural to presume that one finds a linear superposition
of M2-brane states with opposite winding on the two-torus spanned by $\tilde x_3$ and the 
M-theory circle $x^{11}$. The presence of such M2-brane states might help to 
explain how the moduli space of the Spin(7) manifold with vanishing fiber can 
be enhanced to the moduli space of the Calabi-Yau fourfold with vanishing fiber. 
In particular, this requires a complexification of the real Spin(7) moduli space 
to form a K\"ahler manifold.

\section{Summary}

In this work we studied the weak-coupling limit of compactifications of F-theory on Spin(7) manifolds 
that are anti-holomorphic quotients of elliptically fibered Calabi-Yau fourfolds using their M-theory duals. This limit is the natural first step 
towards understanding the physics associated to this class of compactifications. We discussed in detail the following two cases. 
In case $(a)$ the fixed-point loci of the anti-holomorphic involution are real three-dimensional subspaces of 
the base $B_3$ and are one-dimensional subspaces of the fibre. Alternatively, in case $(b)$ the 
fixed-point loci in the base are only one-dimensional. In both cases one of the four macroscopic 
dimensions in F-theory is an interval. We found that the weak-coupling limit of   
case $(a)$ corresponds to a Type IIB compactification with space-time filling O7-planes as well as 
O5-planes and X5-planes localized at the boundary of the interval. 
The X5-planes are objects that have been identified in perturbative string theory 
in the past \cite{Kutasov:1995te,Sen:1996na,Sen:1998rg,Sen:1998ii,Bergman:1998xv,Hellerman:2005ja} 
and correspond to the S-duals of an O5-D5 system. 
In case $(b)$ we found a more complex system of objects consisting of space-time filling 
O7- and O3-planes as well as exotic O3-planes and X3-planes
localised on the boundary of the interval and confined to a six-dimensional orbifold singularity.

We analysed the supersymmetry properties of these configurations and showed that the objects 
present can be mutually supersymmetric in a toroidal setup. For case $(a)$ we have also argued 
that the mutual supersymmetry is possible if the torus is replaced by a Calabi-Yau threefold. 
It would be desirable to establish similar arguments for case $(b)$.
Using our results we were able to argue that for these configurations, on the Type IIB side, the bulk 
preserves four real supercharges while the boundary preserves only two. 
Effective theories with these properties have been studied in \cite{Belyaev:2008xk,Belyaev:2008ex,Howe:2011tm}.
We therefore conclude that in the infinite interval limit supersymmetry is enhanced 
to $\cN=1$ in four dimensions. We argued that this effect can also be understood on the Type IIA side
in terms of string winding modes which become light in the vanishing interval limit. 
The picture that arises in the weak-coupling limit leads to the expectation that, in the absence of additional branes or fluxes, 
this effect persists at strong coupling and supersymmetry is enhanced in general by M2-brane winding states becoming 
light on the M-theory side. This generalisation is a highly non-trivial process of 
supersymmetry enhancement to four supercharges in the singular limit of certain 
Spin(7) manifolds.

Our work is only part of an initial exploration of F-theory dual of M-theory on suitably fibered Spin(7) manifolds. 
This is in principle a rich arena of new string vacua, and we showed that even in the simplest weak-coupling limits 
the resulting constructions are rather unusual supporting, for instance, O7-, O5- and X5-planes simultaneously or exotic O3-planes and X3-planes. 
There are many directions to explore. One of the more immediate open problems is to find an understanding of the case where 
the anti-holomorphic involution acts freely on the fibre rendering it a Klein-bottle. 
It would be interesting to study the objects present in such a vacuum by using, for example, the results of \cite{Aharony:2007du}. 
A more mathematical direction would be to construct explicit examples of Spin(7) manifolds that support the 
different cases of fixed-point loci we have studied. The constructions of Joyce only admit  
fixed points in the geometry that are resolved to obtain a smooth Spin(7) manifold.
However, the method of of quotienting by an involution is more generally applicable
and it would be an interesting challenge to construct the resolved geometries of the 
different cases. A possible guiding principle to achieve this is provided by our 
identification of the weak-coupling objects, such as O6-planes, located at the fixed 
points and their known up-lift into smooth M-theory geometries in the spirit of \cite{Gomis:2001vk}.

From a more phenomenological perspective the fact that these constructions are based on compactifications that, 
at a general point in moduli space, preserve only two supercharges means they potentially could be useful 
for understanding vacua with high-scale supersymmetry breaking in string theory. 
Although we argued that supersymmetry is restored in the simplest cases it is likely that more general 
constructions can be found where the four-dimensional limit preserves no supersymmetry. 
Indeed if supersymmetry were completely broken on the boundary of the interval on the M-theory side, 
for example by fractional branes, it could lead to a scenario where the size of the interval on the F-theory 
side would interpolate between $\cN=0$ and $\cN=1$ four-dimensional supersymmetry.
The non-supersymmetric non-compact limit could be phenomenologically appealing. 

There are a number of further effects that are worth studying within a non-supersymmetric setup. 
For example, an interesting aspect of the X5-planes is that they support non-BPS but stable states \cite{Sen:1998rg,Sen:1998ii,Bergman:1998xv}. 
The stability of the state is guaranteed as it is the lightest state charged under the U(1)
arising from the twisted sector of the X5-plane. It is a particle in Type IIB, similar to a D0-brane in Type IIA, 
which is confined to lie on the X5-plane. Such a state can be thought of as the S-dual to an open string stretching 
between the D5-brane and its orientifold image across the O5-plane.
The ground state of this string is projected out once the D5-brane sits 
on top of the O5-plane, and so the lightest state is an excited oscillator.\footnote{It can also be seen through the 
tachyonic mode of a ${\rm D1}-\overline{\rm D1}$ state \cite{Sen:1998ii}.} It is interesting that such a stable non-supersymmetric 
state arises naturally in such setups. In our setups these non-BPS states are localised 
at the boundaries of the interval,  
and therefore there phenomenological impact is diluted by the interval length. 
However, it is conceivable that in alternative constructions one finds these non-BPS states in 
the bulk such that this dilution does not occur.

More generally Spin(7) compactifications are also interesting from a 
purely three-dimensional perspective in the context of geometric engineering 
of field theories from M-theory \cite{Gukov:2001hf,Curio:2001dz,Acharya:2002vs,Gukov:2002zg,Becker:2003wb,Forcella:2009jj,Bonetti:2013fma}. 
Indeed, the vacua studied in this work are part of a relatively unexplored region of string theory, 
and the potential applications of these constructions are therefore as yet not sharply defined. 
Also much work remains to understand the objects that appear in these geometries and to make 
progress in the even more challenging task of constructing the different geometries explicitly and 
resolving them. Our work provides evidence that F-theory on Spin(7) manifolds 
can be defined and suggests an intriguing decompactification limit.
Its possible relevance to supersymmetry breaking in string theory, makes this an interesting 
field to explore.

\vspace*{.5cm}
\noindent
\subsection*{Acknowledgments}

We would like to thank David Andriot, Ralph Blumenhagen, Emilian Dudas, Mark Goodsell, Jan Keitel, 
Raffaele Savelli, and Cumrun Vafa for interesting discussions
and correspondence. The work of FB, TG and TP was supported by a research grant of the Max Planck Society. The
work of EP is supported by the Heidelberg Graduate School for Fundamental Physics.

\begin{appendix}
\vspace{2cm} 
\noindent {\bf \LARGE Appendix}

\section{Symmetry Algebras and T-Duality} \label{SymAlgAndTDual}

In this work we have described several quotients which are built from a set of fundamental symmetry actions. 
These include $\O_p$ which is the world-sheet parity inversion, $F_L$ which is the left-moving fermion number 
and $R_{m n r s} = R_m R_n R_r R_s$ where $R_m$ describes the parity inversion $x^{m} \ra - x^m$. These satisfy the algebra 
\begin{gather}
\baed
\O_p^2 & = 1\ , &
R_m^2 & = 1\ , &
((-1)^{F_L})^2 & = 1 \ , \nn \\
\O_p (-1)^{F_L} &=  (-1)^{F_R} \O_p \ ,&
\O_p R_m &=  R_m \O_p \ ,&
R_m (-1)^{F_L}  & = (-1)^{F_L} R_m \ ,  \nn \\
\eaed\\
R_m R_n  =  (-1)^{F_L + F_R}  R_n R_m  \quad \text{if} \quad n \neq m \ .
\end{gather}
Defining $R_m$ as a parity inversion implies a definition of the action of $R_m$ on fermions that is only unique up to a phase. 
Here we have made a choice to discuss $R_m^2  = 1 $. This convention is 
appropriate for the way we describe O$p$-planes 
and is consistent with the conventions of \cite{Blumenhagen:2013fgp}.\footnote{Other conventions can lead to $R_m^2 = (-1)^{F_L + F_R}$.}

Under T-duality these transformations have the following properties 
\ba
T_m (-1)^{F_L} T_m^{-1} &= (-1)^{F_L} \ ,& 
T_m \O_p T_m^{-1} &= \O_p R_m \ , \nn \\ 
T_m R_m T_m^{-1} & = R_m \ ,& 
T_m R_n T_m^{-1} & = R_n (-1)^{F_L}   \quad \text{if} \quad n \neq m \ ,
\ea
where $T_m$ represents T-duality in the $m$ direction. 

These actions can then be lifted to symmetries of M-theory as 
\ba
R_m &\ra R_m \ ,& 
(-1)^{F_L} &\ra R_{11} \cC \ ,& 
(-1)^{F_R} &\ra R_{11} \cC \ ,& 
\O &\ra  \cC \ ,
\ea
where $R_{11}$ is the inversion of the M-theory circle and $\cC$ acts on the M-theory three-form as $C_3 \ra - C_3$. 

\end{appendix}



\end{document}